\documentstyle[epsfig]{aipproc}
\def\simless{\hbox{\rlap{$<$}\lower.5ex\hbox{$\sim$}\ }}
\def\simgreat{\hbox{\rlap{$>$}\lower.5ex\hbox{$\sim$}\ }}

\begin{document}

\pagestyle{empty}

%\doublespace

\title{Probing Strong Gravitational Fields in X-ray Novae}

\author{Jeffrey E. McClintock}
\address{Center for Astrophysics \\ Cambridge, MA 02138}

\maketitle

\begin{abstract}
Most X-ray novae (aka soft X-ray transients) contain black hole primaries. 
In particular, the large mass functions measured for six X-ray novae 
directly clinch the argument (within general relativity) that they
contain black holes.  These firm dynamical results are discussed, and
the urgent need to determine precise masses for black holes is stressed.  
The dynamical evidence for black holes is convincing but it is indirect.  
Now it appears that {\it direct} evidence may be at hand. Three recent 
studies have revealed phenomena that very likely probe strong gravitational 
fields: (1) a comparison of the luminosities of black hole systems and
neutron star systems has yielded compelling evidence for the existence
of event horizons; (2) RXTE observations of fast, stable QPOs have probed the
very inner accretion disks of two black holes; and (3) three different 
types of low energy spectra have been linked to different black-hole spin states 
(e.g. Kerr {\it vs.} Schwarzschild).  
\end{abstract}

\section*{Introduction}
A ``classical" X-ray nova typically brightens in X-rays by as much as 
$10^{7}$ in a week and then decays back into quiescence over the course of 
a year.  The optical counterpart, if visible, undergoes a similar outburst 
cycle, although the amplitude is more modest (typically $\simless$~5-7 
mag).  For reviews on the X-ray and optical properties of X-ray novae see 
Tanaka \& Shibizaki \cite{tanaka96} and van Paradijs \& McClintock 
\cite{vanpar95}, and for a  recent compendium of light curves see Chen 
et al. \cite{chen97}.  Not all light curves of X-ray novae have the 
classical shape just described; for example, some are flat-topped or 
triangular in shape \cite{chen97}. A few others cannot be described simply 
and are truly extraordinary, such as the light curves of the superluminal 
radio jet sources GRO J1655-40 and GRS1915+105 \cite{asm97}.

It is remarkable that most X-ray novae contain black holes since black
hole primaries are rare among the persistent X-ray binaries \cite{vanpar95}.
The firm evidence for black holes in X-ray novae is derived from radial 
velocity data, which has been obtained in quiescence for eight systems 
(Table 1). Several additional X-ray novae probably contain black hole 
primaries based on the X-ray spectral/temporal properties they share
with the black hole systems in Table 1, and also with Cygnus~X-1 and 
LMC~X-3; these include, for example, GRS1009-45, GS1354-64, A1524-62, 
4U1630-47, GRS1716-249 and GRS1915+105. Finally, a relatively few X-ray 
novae contain neutron stars: for example, Cen~X-4, 4U1608-52, H1658-298 
and Aql X-1.  In this paper, we restrict ourselves to discussing recent 
results obtained for the eight dynamical black hole binaries (Table 1) plus 
the superluminal source and  microquasar GRS1915+105 \cite{mirabel94}.
\newline

\pagestyle{plain}

\begin{table}[h]
\caption{Selected Dynamical Data for Eight Black-Hole X-ray Novae}
\begin{tabular}{lcccl}
Source & P$_{orb}$(hr)&K (km
s$^{-1}$)&f(M/M$_{\odot}$)=PK$^{3}$/2$\pi$G&Reference \\
\tableline
V404 Cyg&155.3\phantom{.8}&208.5$\pm$0.7&6.08$\pm$0.06&\cite{casares94} \\
GS2000+25&\phantom{15}8.3&518.4$\pm$3.5&4.97$\pm$0.10&\cite{filip95a} \\
XN Oph 1977\tablenote{XN $\equiv$ X-ray Nova}&\phantom{1}
12.5&447.6$\pm$3.9&4.86$\pm$0.13&\cite{filip97}\cite{remillard96}\\
GRO J1655-40&\phantom{1}62.9&228.2$\pm$2.2&3.24$\pm$0.09&\cite{orosz97a} \\
XN Mus 1991$^{\rm a}$&\phantom{1}10.4&406$\pm$7,
420.8$\pm$6.3&3.01$\pm$0.15, 3.34$\pm$0.15&\cite{orosz96,casares97a} \\
A0620-00&\phantom{15}7.8&443$\pm$4, 433$\pm$3&2.91$\pm$0.08,
2.72$\pm$0.06&\cite{orosz97b,marsh94}\\
GRO J0422+32&\phantom{15}5.1&380.6$\pm$6.5&1.21$\pm$0.06&\cite{filip95b} \\
4U1543-47&\phantom{1}27.0&124$\pm$4&0.22$\pm$0.02&\cite{orosz97c}\\
\end{tabular}
\end{table}
 
\section*{The Eight Black Hole X-ray Novae}
The absorption-line velocities of the secondary star can be determined
very precisely in quiescent X-ray novae because the non-stellar light from
the accretion disk and/or the ADAF region (see below) is small or modest 
compared to the light of the secondary. For example, consider the 
semiamplitude of the radial velocity curve, K, for V404 Cyg, A0620-00 and 
XN Mus 1991, which are the three brightest systems listed in Table 1.  For 
V404 Cyg the uncertainty in K is only $\sim$ 0.3\% (see Table 1), based on 
extensive observations made during four observing seasons.  The 
uncertainties are somewhat larger for two other systems: two independent 
determinations of K differ by a few standard deviations for A0620-00 and 
XN Mus 1991, which corresponds to differences of 2.3\% and 3.6\%, 
respectively (see Table~1). Assuming these small differences are significant, 
they may be due to measurement error, or they may be due to systematic effects 
associated with the tidally-distorted secondary \cite{vanpar95}.  In either 
case, the observed errors in K contribute little ($\simless$ 10\%) to the 
uncertainty in the mass of the black hole, $M_{1}$, compared to the 
uncertainties due to the orbital inclination angle, $i$ (see below). 
Uncertainties in the mass of the secondary, $M_{2}$, are generally less 
important than those in $i$.

The value of the mass function is an absolute lower limit on the mass of the
compact primary \cite{vanpar95}:

\begin{equation}
f(M) \equiv {M_{1}^{3}\sin^{3}i \over (M_{1}+M_{2})^{2}} = {PK^{3}
\over 2\pi G}.
\end{equation}

\noindent Thus the compact primaries in the first six systems in Table~1, 
which have large mass functions and  M$_{1} \simgreat 3~M_{\odot}$, are too 
massive to be neutron stars and are almost certainly black holes within 
general relativity \cite{kalogera96}. The last two systems in Table~1, 
GRO J0422+32 and 4U1543-47, have small mass functions. Consequently, they 
are much weaker dynamical black-hole candidates because one must also set 
secure limits on $i$ and/or M$_{2}$, which is generally much less 
straightforward than measuring K. 
  
To determine the mass of the black hole (not just a limit on the mass), one 
must determine both $i$ and M$_{2}$ in addition to the value of the mass 
function.  The inclination angle $i$, which is the more important parameter, 
can be difficult to nail down.  For example, two studies of A0620-00 have 
given the following discrepant results: $62^{\rm o} < i < 71^{\rm o} 
\Rightarrow $  M$_{1} \approx 5~M_{\odot}$ \cite{haswell93} and $31^{\rm o} 
< i <  54^{\rm o} \Rightarrow $ M$_{1} \approx 10~M_{\odot}$ 
\cite{shahbaz94}.  For GRO J0422+32 the discrepancy is even larger: $i = 48 
\pm 3^{\rm o} \Rightarrow M_{1} = 3.6 \pm 0.3~M_{\odot}$ \cite{filip95b}, 
and $i < 31^{\rm o} \Rightarrow M_{1} > 9~M_{\odot}$ \cite{beekman97}. The 
one exceptionally clear-cut case is GRO J1655-40: the high quality of the
four-color ellipsoidal light curve data and the excellent model fit yield a
precise inclination of 69.5 $\pm 0.08^{\rm o}$; the black hole mass is 
$M_{1} = 7.02 \pm 0.22~M_{\odot}$ \cite{orosz97a}.  See also van der Hooft
et al. \cite{hooft98} for a more conservative appraisal. 

Because of the generally large systematic errors just discussed, we choose
not to summarize the masses of the black hole primaries. Instead we refer 
the reader to the papers cited in Table 1 and to a recent compilation by 
Bailyn et al. \cite{bailyn97} who examine the distribution of the masses of 
black holes in X-ray novae.  They argue that these masses are all 
consistent with a single value of $7~M_{\odot}$, {\it except} for V404~Cyg
which has a distinctly higher mass, $M_{1} \sim 12~M_{\odot}$.  In a second 
compilation, Casares \cite{casares95b} compares the masses of 17 neutron 
stars to the lower limits on the masses of the dynamical black holes.

The six compact X-ray sources discussed above with M$_{1} \simgreat 
3~M_{\odot}$ plus the two high-mass X-ray binaries Cyg X-1 and LMC X-3 are 
almost certainly black holes within general relativity.  It would be 
difficult to overstate the importance of this conclusion because it is the 
fundamental ground for all studies of accreting black holes.  The dynamical 
test has proved to be the only acid test for the presence of a black hole; 
a long history of attempts to distinguish black holes from neutron stars 
via spectral and temporal X-ray signatures have failed to yield any 
reliable criteria \cite{tanaka96} with one possible exception 
\cite{barret96}.  As discussed below, new attempts are now being made: 
sensitive imaging observations of the faint quiescent states of accreting black 
holes are providing the first evidence for event horizons, and timing and 
spectral studies with RXTE are revealing probable relativistic effects.  However, 
the underpinning for this new and promising work, which is discussed in 
the following sections, is the dynamical determinations that eight X-ray 
binaries almost certainly contain black holes.

\section*{Early Results from Near the Event Horizon}
As discussed above, the best evidence for black holes comes from dynamical 
studies of X-ray binaries that contain massive compact stars.  
Unfortunately, this evidence alone is not decisive because it presumes that 
general relativity is the correct theory of strong gravity \cite{mcclintock88}. 
Moreover, the dynamical evidence cannot rule out exotic models of massive, collapsed 
stars that have tangible surfaces \cite{bahcall90}. For some time it has 
been clear that to advance further we must attempt to make clean 
quantitative measurements of relativistic effects that occur in the near
vicinity of a black hole or a neutron star.  However, it has been unclear 
how this might be done \cite{mcclintock88}.

Recently, there have been major advances in both theory (e.g. the ADAF model
of accretion; see below) and in observational capabilities (e.g. RXTE).
As a consequence, it appears that we are beginning to discern effects
predicted by general relativity that occur in the near vicinity of 
accreting black holes. We briefly discuss progress on three fronts: (1) 
evidence for the existence of event horizons; (2) fast, stable QPOs that 
emanate from near the event horizon; and (3) three basic types of low 
energy spectra that may reveal the metric of the black hole.

\subsection*{The ADAF Model and Evidence for Event Horizons}
An advection-dominated accretion flow (ADAF) is defined as one in which
a large fraction of the viscously generated heat is advected with the 
accreting gas and only a small fraction of the energy is radiated. The 
ADAF model has been applied very successfully in fitting the broadband
spectra (optical to X-ray) of X-ray novae in quiescence.  In the following, 
to be definite, we refer the reader to a discussion of the model for 
V404 Cyg \cite{narayan97a}, the brightest X-ray nova in quiescence. The low 
accretion rate flow proceeds via a standard thin accretion disk, which 
transforms itself by evaporation totally into a corona or ADAF at 
$R_{\rm tr} \sim 10^{4} R_{S}$ ($R_{S}$ is the Schwarzschild radius).  The 
hot flow in the inner region is quasi-spherical and optically thin.  When 
the gas reaches the black hole, the ion and electron temperatures are 
$T_{i} \sim 10^{12.5}K$ and  $T_{e} \sim 10^{9.5}K$. The bulk of the X-ray 
and optical emission is produced within $R_{em} \sim 10 R_{S} \sim$ 300 km
of the black hole.  

The radiation from the ADAF is in the form of Comptonized synchrotron and 
bremsstrahlung emission and has a broad spectrum extending from the near-IR 
to soft gamma-rays. The X-ray spectrum and non-stellar optical spectrum for 
V404 Cyg are very well fitted by this model.  In addition to $R_{\rm tr}$, 
the model has five parameters; however, four of these have very little 
effect on the computed spectrum (see Figures 3-5 in Narayan et al. 
\cite{narayan97a}). The fifth parameter, the mass transfer rate, was 
determined by normalizing the model to the data at 3.5 keV: \.{M} $\sim 
1 \times 10^{-9} M_{\odot}$ yr$^{-1}$. The overall radiative efficiency of 
the accretion flow  is $\sim$~0.05\% (compared to $\sim$~10\% for standard 
disk accretion) \cite{narayan97a}.

When a hot ADAF flow with its extremely low radiative efficiency encounters
a quiescent black hole, then the enormous thermal energy stored in the gas 
($T_{i} \sim 10^{12.5}K$) simply disappears through the event horizon.  The 
energy flow rate into the black hole is $\sim 0.1 \times$ 
\.{M}${\rm c}^{2}$.

What about a neutron star accretor?  It is reasonable to expect that a
quiescent X-ray nova with a neutron star primary (e.g. Cen X-4) will also
accrete via an ADAF with a low radiative efficiency ($\sim$~0.05\%).  Now,
however, the energy of the superheated gas cannot disappear.  The gas falls
on the neutron star and heats its surface.  Once a steady state is reached,
the star's luminosity will be the same as for a thin disk and the accretion
efficiency will be $\sim$~10\%, far higher than for the black hole.

\begin{figure}[h]
\centerline{\epsfig{file=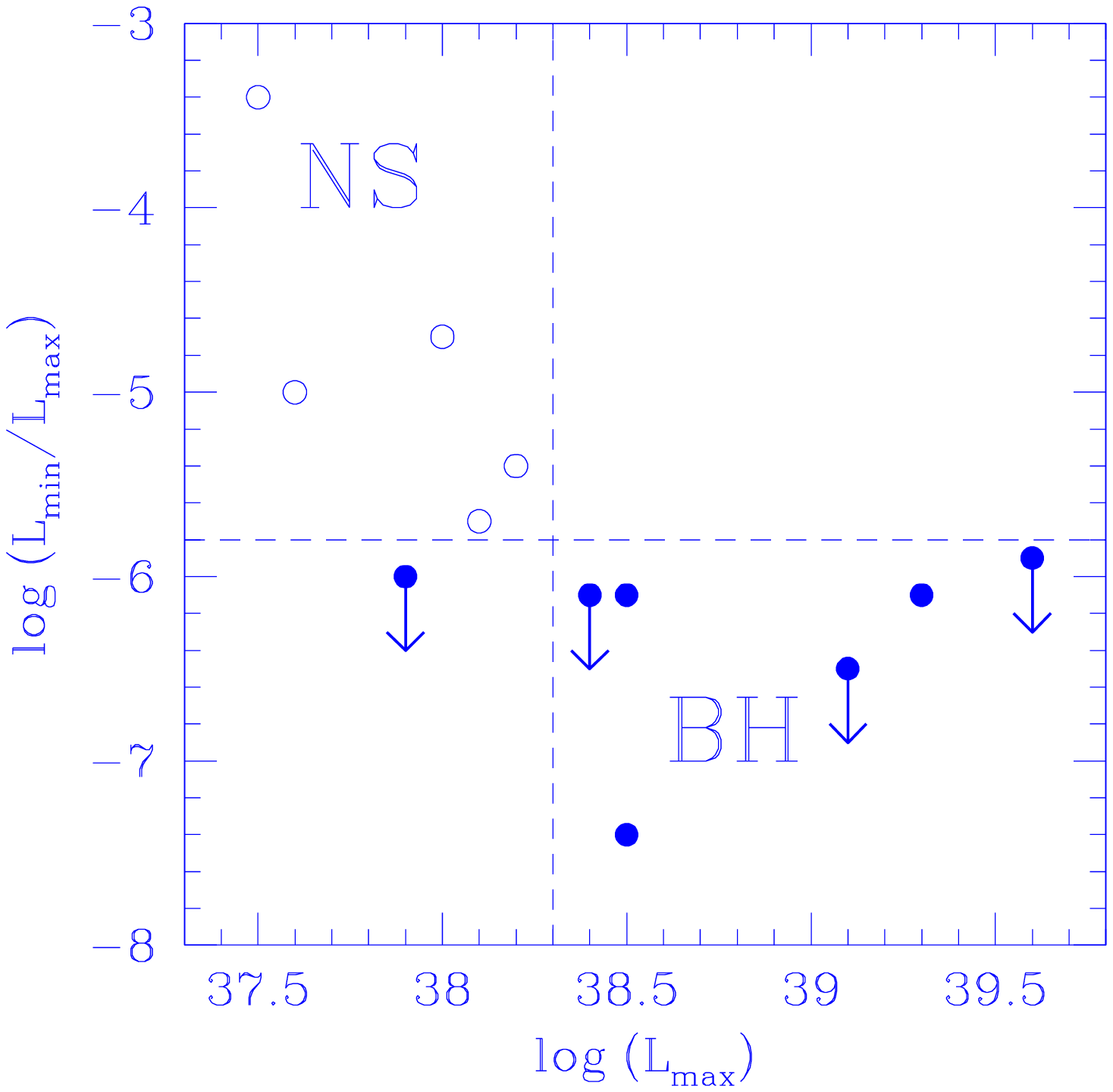,height=3.6in,width=3.6in}}
%\centerline{\epsfig{file=fig.ps}}
\vspace{10pt}
\caption{}
\end{figure}

Thus, for the same \.{M} in quiescence we expect a black hole to be
substantially less luminous than a neutron star, precisely as observed 
\cite{narayan97b}. Figure 1 compares $L_{min}/L_{max}$ for five neutron
stars (open circles) to this same quantity for all of the black holes
listed in Table~1 except XN Oph 1977 (filled circles). 
Without exception, every black-hole X-ray 
nova in quiescence has a smaller value of $L_{min}/L_{max}$ than any of the 
quiescent neutron-star systems.  In fact, even if one treats the limits as
data points, the luminosity swing from outburst to quiescence is higher for the 
black holes by a factor of $\sim$100 on average.  Furthermore, the 
difference in luminosity swings between black holes and neutron stars is 
statistically significant at the $\sim$99.5\% level of confidence 
\cite{garcia97}. The most plausible explanation for this difference is that 
black holes have event horizons and neutron stars do not.  Thus a black hole 
will hide any thermal energy it accretes, whereas a neutron star must 
re-radiate any thermal energy it accretes.  For further discussion of this 
evidence for black-hole event horizons and a thorough discussion of the data 
plotted in Figure~1, see Garcia et al. \cite{garcia97}. 

In our black hole sample we have included only the dynamically confirmed 
black holes (Table 1) and we have considered only data of very high 
quality collected by imaging telescopes (Einstein, ROSAT and ASCA). On 
the other hand, Chen et al. \cite{chen97b,chen97} have argued that in this 
comparison one should include the data for all plausible quiescent X-ray 
novae, even if the data quality is relatively poor and the nature of the 
compact primary is uncertain.  Accordingly, Chen et al. \cite{chen97b} have 
presented an alternative version of Figure 1 above. The upper half of their
figure is populated with an additional dozen or so upper limits and 
detections of probable neutron stars.  On casual inspection these relatively
low sensitivity data appear to blur the difference between the luminosity 
swings of black holes and neutron stars.  In fact, however, a proper statistical
analysis of all the data would very likely show that the addition of these low 
quality data does not affect the significance of the results quoted above.

Indeed, data of low qualilty are irrelevant: what is needed for progress 
are data of higher quality.  Note that only upper limits are available for 
four of the seven black hole systems in quiescence (whereas all of the 
neutron stars are detected).  It is probable that all these black holes 
will be detected soon by AXAF or XMM, which can only serve to make the 
difference in luminosity swing between black holes and neutron stars more 
dramatic.

\subsection*{Black Hole Spin via Fast Quasiperiodic Oscillations}
High frequency QPOs have been observed with RXTE from two black hole 
candidates, the microquasars GRS1915+105 and GRO J1655-40.  Most reliable 
is GRS1915+105, which very frequently shows a sharp (Q $\approx$ 20) QPO 
at a frequency of 67 Hz \cite{morgan97}.  The variation in the central 
frequency is only 1 Hz.  The RMS amplitude is typically about 1\%; however, 
both the amplitude and the phase lag of the QPO increase quite 
significantly with energy.

GRO J1655-40 has displayed a high frequency QPO at 300 Hz during seven
observations when the spectrum was exceptionally hard.
Combining all these data, the feature has a total statistical significance
of 14$\sigma$, a Q of $\approx$ 3, and an amplitude near 0.8\% 
\cite{remillard97}. 

Several models for the black hole QPOs have been discussed.  The simplest
of these relates the observed frequency to the frequency of the minimum
stable orbit.  For a Schwarzschild black hole, this picture implies masses 
of 33~M$_{\odot}$ for GRS1915+105 \cite{morgan97} and 7.4~M$_{\odot}$ for 
GRO J1655-40 \cite{remillard97}.  The latter case is most intriguing 
because the dynamically-determined mass for GRO J1655-40 is $7.02 \pm 
0.22~M_{\odot}$ \cite{orosz97a}!  The ``diskoseismic" model invokes radial 
disk modes with eigenfrequencies that depend on the mass and angular 
momentum of the black hole.  For GRS1915+105, this model predicts a 
10.6~M$_{\odot}$ Schwarzschild and a 36.3~M$_{\odot}$ extreme-Kerr black 
hole \cite{nowak97}.  It has also been suggested that the fast QPOs are 
due to the precession frequency of an accretion disk that is experiencing 
frame dragging caused by a rotating black hole \cite{cui97}.  In this 
model, the observed QPO provides a direct measure of the black hole's 
angular momentum once its mass is determined.

Presently it is unclear which model, if any, correctly describes the 
observed QPOs.  Nevertheless, it is plain that these QPOs are effective 
probes of the very inner regions of the accretion flow.  

\subsection*{Black Hole Spin via the Low Energy Spectrum}
An ultrasoft spectrum---i.e. one that is distinctly softer than the spectra 
of neutron star sources---has long been recognized as a useful signature 
of a black hole \cite{white84,tanaka96}.   An ultrasoft component in the 
spectrum of a luminous black hole is well explained by quasi-blackbody 
emission from the inner portion of an accretion disk, which extends down to 
the minimum stable orbit.  The ultrasoft spectral component is not, however, 
an infallible signature of a black hole because it is sometimes absent. 
For example, see Figure 2 in Sunyaev et al. \cite{sunyaev93}, which shows 
the high-luminosity spectra of three of the black holes listed in Table 1: 
the ultrasoft component is completely absent in both V404 Cyg (the 
most secure black hole candidate) and in GRO J0422+32, whereas it is strong 
in GS2000+25. 

Zhang et al. \cite{zhang97} argue that this extraordinary spectral 
difference, and also the harder low-energy spectra of the microquasars, 
GRS1915+105 and GRO J1655-40, are due to the rotation of the black hole 
relative to its accretion disk.  Specifically, they argue that (1) the 
harder spectra of the microquasars is due to the near-extreme prograde 
rotation of their Kerr black holes, (2) the systems with ultrasoft spectral 
components (e.g. GS2000+25 and XN Mus 1991, etc.) contain slowly rotating 
Schwarzschild black holes, and (3) the systems with no ultrasoft component 
(e.g. V404 Cyg, GRO J0422+32, etc.) contain near-extreme retrograde black 
holes.  The authors consider a number of subtelties in deriving relations 
between the blackbody luminosity and disk temperature {\it vs.} the angular
momentum of the black hole.  But qualitatively, their results can be 
easily understood in terms of the depth of the gravitational potential 
at the location of the minumum stable orbit---i.e. at the inner edge of
the disk. For extreme Kerr, Schwarzschild, and extreme retrograde holes, 
the radii of the minimum stable orbits are M, 6M and 9M (c = G = 1),  
and the approximate accretion efficiencies are 30\%, 6\% and 3\%, 
respectively.

The most compelling example considered by Zhang et al. is GRO J1655-40 
because its black hole mass, distance and other parameters are 
well determined.  Applying their model, they conclude that the primary is
a Kerr black hole spinning at between 70\% and 100\% of the maximum
rate.  Unfortunately, the other black hole systems provide few additional
constraints on the model of Zhang et al.  The good result for 
GRO J1655-40 underscores the need for more work on determining inclinations,
masses and distances for selected X-ray novae.

\section*{Conclusions}

It must be considered good fortune that X-ray novae have revealed fast,
stable QPOs that almost certainly come from the near vicinity of the 
black hole primary.  It is also fortunate that RXTE has the collecting area 
and telemetry capability to scrutinize these relatively weak temporal 
features.  Indeed, as a consequence of the `no-hair' theorem, many of us 
have occasionally feared that it might prove impossible to achieve a useful 
quantitative test of general relativity by observing an accreting black 
hole.  Unlike the case of a magnetized neutron star, the erratic 
magnetohydrodynamic effects of gas orbiting a black hole might hopelessly 
confuse any relativistic effects \cite{blandford87}.  It now seems that the 
prospects are bright for tests of general relativity in the strongest 
possible gravitational fields. 

In addition to the QPOs, it appears that the low-energy spectra of luminous X-ray 
novae, which almost certainly originate in strong fields near the minimum 
stable orbit, have yielded important information on the metric.  Finally, 
the dramatic difference in luminosity between a hot ADAF flow that impinges 
on a black hole and one that strikes the surface of a neutron star provides 
the first evidence for the intangible entity that defines a black hole, 
namely, its event horizon.

The foundation for these studies has been the firm evidence that a 
half-dozen X-ray novae contain black holes (Table 1).  The discovery of 
additional dynamically-determined black holes remains an important task.  
However, much future dynamical work needs to be directed toward determining 
the precise masses of selected black holes.  Presently, only GRO J1655-40 
has both a precise mass and a good distance.  Consequently, its black hole 
served as a touchstone for interpreting the fast QPOs and the low-energy 
spectral data.

\end{document}